\documentclass[11pt]{JHEP3} 

\usepackage{epsfig,multicol,bbm}

\newcommand\fverb{\setbox\fverbbox=\hbox\bgroup\verb}
\newcommand\fverbdo{\egroup\medskip\noindent%
            \fbox{\unhbox\fverbbox}\ }
\newcommand\fverbit{\egroup\item[\fbox{\unhbox\fverbbox}]}
\newbox\fverbbox

\usepackage{amsmath}
\usepackage{graphicx}
\usepackage{amssymb}
\usepackage{latexsym}
\usepackage{graphicx}
\usepackage{dcolumn}
\usepackage{bm}

\def\be{\begin{equation}}
\def\ee{\end{equation}}

\def\a{\alpha}

\def\df{\dot{\phi}}

\def\la{\langle}
\def\ra{\rangle}
\def\vk{\textbf{k}}

\title{Primordial Non-Gaussianities of General Multiple Field
Inflation}

\author{Xian Gao\\
    Institute of Theoretical Physics, Chinese Academy of Sciences,\\
    No.55, ZhongGuan Cun East Road, HaiDian District, Beijing 100080, China\\
    E-mail: \email{gaoxian@itp.ac.cn}}

\abstract{We perform a general study of the primordial scalar
non-Gaussianities in multi-field inflationary models in Einstein
gravity. We consider models governed by a Lagrangian which is a
general function of the scalar fields and their first-order
spacetime derivatives. We use $\delta N$ formalism to relate scalar
fields and the curvature perturbations. We calculate the explicit
cubic-order perturbative action and the three-point function of
curvature perturbation evaluated at the horizon-crossing. Under
reasonable assumptions, in the small slow-varying parameters limit
and with a sound speed $c_s$ close to one, we find that the
non-Gaussianity is completely determined by these slow-varying
parameters and some other parameters which are determined by the
structure of the inflationary model. Our work generalizes previous
results, and implies the possibility of the existence of large
non-Gaussianity in model constructing, and it would be also useful
to study the non-Gaussianity in multi-field inflationary models
which will be constructed in the future.}

\begin{document}



\section{Introduction}

Inflation has been a very successful paradigm for understanding the
evolution of the very early universe \cite{inflation_bible}. It not
only naturally provides a way to solve flatness and horizon
problems, but also generates density perturbations as seeds for the
large-scale structure in the universe. Inflation is most commonly
discussed in terms of a potential energy which is a function of a
single, slowly rolling scalar field. Such models generically predict
an almost scale-invariant spectrum and an almost Gaussian
distribution of adiabatic density perturbations on super-Hubble
scales \cite{pertubation_bible}. These generic predictions are
consistent with recent cosmological observations of the temperature
anisotropy and polarization of the Cosmic Microwave Background
\cite{wmap5}.

The measurement of primordial perturbations provides increasingly
precise determination of the spectrum index $n_s$ and
tensor-to-scalar ratio $r$, which come from the power spectrum, in
other words, the two-point correlation function of the curvature
perturbation. However, there are many alternative models of
inflation are able to give the similar predictions, and thus there
is still considerable ambiguity in constructing the real
inflationary model.

In contrast, the non-Gaussian component of the scalar perturbations
is characterized by the correlation functions beyond two-point, e.g.
the three-point function of the fluctuations, which is a nontrivial
function of three variables and will provide us more information
beyond the power spectrum. Furthermore, as described above, the
non-Gaussianity of distribution of primordial fluctuations predicted
by the simplest model of inflation is well below the current limit
of measurement \cite{0210603,0209156,9208001,9312033}. Therefore,
any detection of large non-Gaussianity would be a significant
challenge to our current understanding of the early universe.

Indeed, there is the possibility of the presence of large
non-Gaussianity with $f_{\mathrm{NL}} \gg 1$ \cite{wmap5,0803.2157}.
The latest observational bound on the three-point function of the
primordial curvature perturbation $\zeta$ from the WMAP 5-year data
states that the local non-gaussianity parameter $f_{\mathrm{NL}}$ is
limited to the values $-9 < f^{\mathrm{local}}_{\mathrm{NL}} < 111$
\cite{wmap5}. In the next few years, with improved experiments like
the Planck satellite, we will measure the CMB anisotropies to an
incredible resolution at $|f_{\mathrm{NL}}| < 5$.

On the theoretical side, the level of non-Gaussianities can be
generally calculated analytically in typical single or
multiple-field inflationary models (see \cite{0406398} for a
review). The three-point function for the standard single-field
inflation models with a canonical kinetic term was performed in an
elegant and gauge invariant way in \cite{0210603,0209156}, and the
corresponding $f_{\mathrm{NL}}$ is $\mathcal{O}(10^{-2})$ and too
small to be detected. In order to find possibly detectable large
non-Gaussianities in single-field cases, models with non-standard
actions have also been studied
\cite{0306122,0411773,0503692,0605045,wangyi}.

In the past few years, intensive efforts have been devoted to
connecting string theory and inflation
\cite{0710.2951,0708.2865,0702059,0612129,0610221}. Indeed, in
models descending from the low-energy limit of superstring theory or
string compactification, there are many light moduli fields
describing the higher-dimensional degrees of freedom, which could
play a role during inflation. Therefore, it is interesting and of
physical significance to generalize the above analysis of
non-Gaussianities in single-field models to the multiple-field
inflation models
\cite{Bento:all,Sakharov:all,Enqvist:all,0207295,Rigopoulos:all,Lyth:all,0504045,0506056,Wands:all,0610296,0506532,0510709,0512368,Barnaby:all,0701247,Yokoyama:all,0709.2666,0709.3299,0711.0760}.
Most of the previous works focus on the models with standard
canonical kinetic terms for the scalar fields involved. It is shown
that in models satisfying the slow-roll conditions,
$f_{\mathrm{NL}}$ is also of order slow-roll parameters as in the
standard single-field case. Models with non-canonical kinetic terms
have also been investigated, in particular the effective
multiple-field DBI-inflation \cite{0709.2666,0709.3299,0711.0760}.
Models with non-trivial field space metric $G_{IJ}$ have also been
considered \cite{0510709}.

In these inflationary models inspired by string theory or other
theory beyond the Standard Model, there are several significant
features different from those in the standard single field models,
where the Lagrangian is of the form $\mathcal{L} =
\sqrt{-g}\left[-\frac{1}{2}(\partial\phi)^2 -V(\phi)\right]
=\sqrt{-g}(X-V)$. Firstly, the scalar field potential is not
necessarily the only degree of freedom in model building, in
contrast to that in the standard single field models. Indeed, in
these string inspired models, it is generally expected that
deviation from the standard kinetic term $X \sim (\partial\phi)^2$
of the scalar field action would arise \cite{0208123}. One would
generally expect loop corrections in the quantum theory to generate
operators in the Lagrangian that are proportional to higher-order
derivatives $X^2$, $X^4$, and so on. Indeed, if the energy scale of
renormalization is of order Planck scale $M_{\mathrm{pl}}$, such
higher-order interactions would be suppressed and be negligible, and
thus the canonical kinetic term would give a good approximation. On
the other hand, if the inflation scale is close to the GUT scale,
these ultraviolet corrections might be significant and of
considerable relevance \cite{k-inflation}. Such models with
non-canonical kinetic terms have been considered previously by a
number of authors
\cite{k-inflation,0101225,0206100,0312099,0409130}. For our purpose,
the deviation from a canonical Lagrangian would be an important
source of possible large non-Gaussianities.

Secondly, the presence of many scalar fields during inflation would
affect the generation of primordial perturbations. In single field
inflation, it is well-known that the curvature perturbation is
conserved on super-horizon scales. However, in multi-field inflation
models, the presence of multiple light fields will lead to the
generation of non-adiabatic field perturbations during inflation.
The curvature perturbation in multi-field inflation can generally
evolve after Hubble exiting, due to the presence of entropy
perturbations, which can be `sources' of curvature perturbation.
This non-trivial evolution of the overall curvature perturbation
would be another source of detectable non-Gaussianity. A general
description of calculating the three-point functions of the
fluctuations in multi-field models was presented recently in
\cite{0504045}, which emphasized this super-horizon evolution of
curvature perturbation. It is also argued in \cite{0506056} that in
some cases, for example in the curvaton scenario
\cite{wands:curvaton,0309033,0110096,0109214,0602285,0607627,0801.0467,0803.2637},
there is the possibility that the large-scale superhorizon-effected
components would dominate the primordial non-Gaussianity, rather
than the components from the microscopic fluctuations. However, in
the absence of any estimation for the microphysical contributions to
the three-point functions, the above analysis is incomplete, and
thus a precise investigation of non-Gaussianities generated from the
fluctuations at the horizon-crossing is needed.

In this work, we consider a very large class of multiple-field
inflation models, in which the dynamics of scalar fields is
described by a Lagrangian $P(X,\phi^I)$, where $P$ is an arbitrary
function of $\mathcal{N}$ scalar fields and their kinetic term
$X=-\frac{1}{2}(\nabla \phi^I)^2$ \cite{0801.1085}. This form of
action includes the standard choice $P=X-V$ as a special case, and
can be viewed as a generalization of the Lagrangian of k-inflation
\cite{k-inflation} to the cases of multiple scalar fields.
Low-energy limit of string theory may also lead to such an action,
such as Dirac-Born-Infeld action \cite{0310221,0402075}, and its
multi-field extension studied recently in
\cite{0709.2666,0709.3299}. In general, one may consider theories
including arbitrary number of higher-order derivatives in the
action, such as $P(\phi,\partial\phi,\partial^2\phi,\cdots)$.
However, if the energy scale of inflation process is much lower than
$M_{\mathrm{pl}}$, the contributions from these higher-order
derivatives will be suppressed and thus can be neglected. Therefore,
we consider models where $P$ contains only $X$. Due to the same
reason, we consider Einstein gravity.

We assume that these scalar fields generate the density
perturbations. We expand the general multi-field Lagrangian to cubic
order of the perturbations $\delta \phi^I$, and use $\delta N$
formalism \cite{deltaN_starobinsky,9507001,9801017,0411220} to
relate curvature perturbation $\zeta$ and these multiple scalar
perturbations $\delta \phi^I$. We calculate the scalar three-point
function of the curvature perturbations following Maldacena et. al.
\cite{0210603}. To control the calculation, we define some small
slow-varying parameters, which can be viewed as the generalization
of the standard slow-roll parameters. In our formalism, although the
Lagrangian is very general, we assume the effective speed of sound
$c_s$ is almost close to one. In the limit of small slow-varying
parameters, under some reasonable assumptions, we finally find that
the non-Gaussianity in these models is completely characterized by
the slow-varying parameters and some other parameters determined by
the concrete structure of $P(X,\phi^I)$. In particular, our result
shows the possibility of the presence of large non-Gaussianities,
due to the deviation from the canonical Lagrangian.

The paper is organized as follows. In the next section, we setup the
general multi-field inflationary models, and derive the background
equations of motion. In order to control the calculation and
analysis the solution, we define some slow-varying parameters in
this general context. Then we give a brief review of calculating
non-Gaussianities during multi-field inflation based on $\delta N$
formalism. In section 3, we develop the second-order theory for the
linear perturbations. We estimate the power spectrum in the limit of
small slow-varying parameters. In section 4, we calculate the exact
cubic perturbation action for this most general multi-field
inflationary model. In section 5, we perform a general calculation
of the three-point function, which represents the central result of
this work. Finally, we make a conclusion in section 6.

We work in natural units, where $c=\hbar=M_{\mathrm{pl}}=1$, and
$M_{\mathrm{pl}}\equiv(8\pi G)^{-1/2}$ is the reduced Planck mass.


\section{Setup}

\subsection{Background}

We consider a large class of multi-field inflation models, which are
constructed from a generic set of $\mathcal{N}$ scalar fields
$\{\phi^I, I=1,2,\cdots,\mathcal{N}\}$ coupled to Einstein gravity.
The action takes the form: \be{\label{action}}
    S = \int d^4 x \sqrt{-g}\left[\frac{1}{2}R +
    P(X,\phi^I)\right]\;,
\ee with kinetic term
\begin{equation}
    X =
    -\frac{1}{2}G_{IJ}g^{\mu\nu}\partial_{\mu}\phi^I\partial_{\nu}\phi^J\;,
\end{equation}
where $g_{\mu\nu}$ is the spacetime metric with signature $(-,+++)$
and $G_{IJ}=G_{IJ}(\phi)$ is the $\mathcal{N}$-dimensional field
space metric, and $P$ is an arbitrary function of $X$ and $\phi$'s.

The energy-momentum tensor of the scalar fields takes the form
\begin{equation}
    T^{\mu\nu} = Pg^{\mu\nu} +
    P_{,X}G_{IJ}\partial^{\mu}\phi^I\partial^{\nu}\phi^J\;,
\end{equation}
where $P_{,X}$ denotes the partial derivative of $P$ with respect to
$X$. In order to consider the background (unperturbed) dynamics, we
suppose that the universe is homogeneous, with a flat
Friedmann-Robertson-Walker metric
\begin{equation}
    ds^2 = -dt^2 + a^2(t)dx^idx^i\;,
\end{equation}
where $a(t)$ is the scale factor and $H=\dot{a}/a$ is the Hubble
parameter. Under this assumption, the energy-momentum tensor of the
scalar fields reduces to that of a perfect fluid, with energy
density
\begin{equation}
    \rho = 2XP_{,X}-P\;,
\end{equation}
and pressure $P = P(X,\phi^I)$.

The equations of motion for the scalar fields derived from
(\ref{action}) are \cite{0801.1085}
\begin{equation}{\label{scalar_eom}}
    \ddot{\phi}^I + \Gamma^I_{JK} \dot{\phi}^J\dot{\phi}^K +
    \left( 3H + \frac{\dot{P_{,X}}}{P_{,X}} \right) \dot{\phi}^I -
    \frac{G^{IJ}}{P_{,X}}P_{,J}=0\;,
\end{equation}
where $P_{,I}$ denotes the derivative of $P$ with respect to
$\phi^I$, and $\Gamma^I_{JK}$ is the Christoffel symbols associated
with the field space metric $G_{IJ}$. The equations of motion of the
gravitational dynamics are Friedmann equation
\begin{equation}
    H^2 = \frac{\rho}{3} \equiv \frac{1}{3}(2XP_{,X} - P)\;,
\end{equation}
and the continuity equation
\begin{equation}
    \dot{\rho} = -3H(\rho + P) \equiv -6H X P_{,X}\;.
\end{equation}
The combination of the above two equations gives another useful
equation \be
    \dot{H} = -XP_{,X}\;.
    \ee

It is convenient to define an effective speed of sound $c_s$
\cite{k-inflation}, as
\begin{equation}{\label{cs}}
    c^2_s \equiv \frac{P_{,X}}{\rho_{,X}} =
    \frac{P_{,X}}{P_{,X}+2XP_{,XX}}\;.
\end{equation}
Note that in models with canonical kinetic terms $P=X-V$, $c_s=1$.

In general one may consider models with an arbitrary field space
metric $G_{IJ}(\phi^I)$. In this work, we focus on the case where
$G_{IJ} =\delta_{IJ}$ and thus $\Gamma^{I}_{JK}=0$, i.e. the field
space is flat. This choice of action already covers a large class of
multi-field inflationary models with non-canonical kinetic terms.
The linear perturbations of models with an arbitrary metric $G_{IJ}$
have been investigated in \cite{0801.1085}.

\subsection{Slow-varying parameters}{\label{section_slow_para}}
For general function $P(X,\phi^I)$, it is difficult to solve the
equations of motion for the scalars (\ref{scalar_eom}) analytically.
In order to capture the main physical picture and investigate the
evolution of the system, the idea is to define some small parameters
to control the dynamics, and to find solutions perturbatively in
power expansions of these small parameters.

In standard single field slow-roll inflation, this condition is
achieved by assuming that the inflaton $\phi$ is rolling slowly in
comparison with the expansion rate $|\df| \ll H$. Similarly, in
general multi-field inflation models, it will prove convenient to
introduce a dimensionless ``slow-varying'' matrix as \cite{0506056}
\begin{equation}
    \epsilon^{IJ} = \frac{P_{,X}\dot{\phi}^I\dot{\phi}^I}{2H^2} = \epsilon^I\epsilon^J\;,
\end{equation}
where \be
    \epsilon^I = \sqrt{\frac{P_{,X}}{2}}\frac{\df^I}{H}\;.
    \ee
When there is only one inflaton field involved, $\epsilon^{IJ}$
reduces to the standard single field slow-roll parameter $\epsilon =
-\dot{H}/H^2$, which can also be expressed as \be
   \epsilon = \mathrm{tr}\epsilon^{IJ} = G_{IJ}\epsilon^{IJ} = -\frac{\dot{H}}{H^2}.
   \ee
It proves useful to decompose $\epsilon$ into two new parameters
$\epsilon_X$ and $\epsilon_{\phi}$ \cite{0503692}, which measure how
the Hubble parameter $H$ varies with the kinetic and potential parts
of scalar fields $\phi^I$ respectively \be
    \epsilon = \epsilon_X + \epsilon_{\phi} =
    -\frac{H_{,X}}{H^2}\dot{X} - \frac{H_{,I}}{H^2}\df^I\;.
    \ee

As in the single field models where we may define $\eta =
\frac{\dot{\epsilon}}{{H\epsilon}}$, here we define another
slow-varying matrix $\eta^{IJ}$ as \be
    \eta^{IJ} \equiv \frac{\dot{\epsilon}^{IJ}}{\epsilon H}\;,
    \ee
which can be written explicitly \be
    \eta^{IJ} = 2\epsilon^{IJ} - \frac{P_{,X}(\ddot{\phi}^I\df^J+\df^I\ddot{\phi}^J) + \dot{P_{,X}}\df^I\df^J}{2H\dot{H}}\;.
    \ee
The matrix $\eta^{IJ}$ generalizes the slow-roll parameter $\eta$ in
single-field inflation models. Note that with a flat target space
metric $G_{IJ}$, we have $\eta = \mathrm{tr}\eta^{IJ} =
\dot{\epsilon}/\epsilon H$ as expected.

Due to the generality of the fuction $P(X,\phi^I)$ and the
complexity of the scalar equations of motion (\ref{scalar_eom}), the
relations between $P_{,I}$ and the rolling of scalar fields
$\dot{\phi}^I$ and $\ddot{\phi}^I$ are complicated. And thus we may
introduce another set of parameters defined as \be
    \tilde{\epsilon}_I = -\frac{P_{,I}}{3\sqrt{2P_{,X}}H^2}\;,
    \ee
which can be viewed as the analogue of the standard slow-roll
parameter of the form $\epsilon =\frac{1}{2} (V'/V)^2$. In models
with standard canonical kinetic terms, it is easy to show that
$|\tilde{\epsilon}^I| \approx |\epsilon^I|$ as expected, and thus
two slow-roll parameters $\epsilon$ and $\eta$ are enough to control
the theory. However, for general function $P$, there is no simple
relation between these two parameters. In order to proceed, we
expect that both $\tilde{\epsilon}^I$ and $\epsilon^I$ are of the
same order, and assume that $\dot{P_{,X}}/(HP_{,X})$ can be
negligible. Therefore, from the scalar equations of motion
(\ref{scalar_eom}), it is easy to see that $\tilde{\epsilon}^I
\approx -\epsilon^I$ as expected. Similarly another matrix is
defined as \be
    \tilde{\eta}_{IJ} = -\frac{P_{,IJ}}{3H^2P_{,X}}\;,
    \ee
this is the analogue of $\eta = V''/V \approx V''/3H^2$ in the
single field case.

Furthermore, we define the dimensionless parameters \be
    u=\frac{1}{c_s^2} -1\,,\qquad s = \frac{\dot{c}_s}{c_s H}\;,
    \ee
where $u$ measures the deviation of the effective sound speed $c_s$
from unity, and $s$ measures the change speed of $c_s$. In models
with canonical kinetic terms, $u=s=0$.

These parameters generalize the usual slow-roll parameters, and in
general depend on the kinetic terms as well as the potential terms.
For generic theories we expect that $|\epsilon|,|u|,|s| \ll 1$
and{\footnote{In this work we consider the case with $c_s$ is very
close to one by assuming $u \sim \mathcal{O}(\epsilon)$. In general,
one may consider models with an arbitrary $c_s$, however, as
addressed in \cite{0709.2666,0709.3299} where the multiple DBI
inflation is investigated, the multi-field effect is suppressed in
the limit of $c_s \ll 1$, which may be a possible source of large
non-Gaussianity in single field models. We will leave this for a
future investigation.}} \be
    \epsilon^{IJ},\eta^{IJ},\tilde{\eta}^{IJ} \sim \mathcal{O}\left(
    {\epsilon}/{\mathcal{N}}\right)\;,
    \ee
and thus $\epsilon^I \sim \tilde{\epsilon}^I \sim
\mathcal{O}\left({\sqrt{{\epsilon}/{\mathcal{N}}}}\right)$, here
$\mathcal{N}$ is the number of the scalar fields. Furthermore, for
those models with non-vanishing ``cross'' derivatives, i.e.
$P_{,XI},P_{,XXI},P_{,XIJ} \neq 0$, we assume that the
`$X$'-derivatives of these slow-varying parameters also satisfy some
smallness relations \be
    \tilde{\epsilon}^I_{,X} \sim
    \frac{P_{,X}}{H^2}{\tilde{\epsilon}}^I\,,\qquad
    \tilde{\epsilon}^I_{,XX} \sim \frac{P_{,X}^2}{H^4}
    \tilde{\epsilon}^I\,, \qquad \tilde{\eta}^{IJ}_{,X} \sim
    \frac{P_{,X}}{H^2}\tilde{\eta}^{IJ}\;.
    \ee
In general, the validity of these conditions depends on the explicit
forms of the models. In this work, we do not try to find explicit
models which satisfy these slow-varying conditions. It is also
addressed in \cite{0503692,0605045} and etc., in the presence of a
non-canonical kinetic term, the smallness of these slow-varying
parameters does not imply that the inflation itself is slow-rolling.

In order to simplify some of the derivations that follows, it will
prove useful to define two parameters which are combinations of
derivatives of $P$ with respect to the kinetic term $X$
{\footnote{From (\ref{cs}),(\ref{Sigma})-(\ref{lambda}), we can
extract some useful relations for later convenience
\[
    XP_{,X} = \Sigma c_s^2\,,\qquad
    X^2P_{,XX} = \Sigma(1-c_s^2)/2\,,\qquad
    X^3P_{,XXX} = 3\lambda/2 - 3\Sigma(1-c_s^2)/4\,,\qquad \frac{1}{c_s^2}-1 = 2X\frac{P_{,XX}}{P_{,X}}.
\] }}
\begin{equation}{\label{Sigma}}
    \Sigma = XP_{,X}+2X^2P_{,XX} = \frac{\epsilon H^2}{c_s^2}\;,
\end{equation}
\begin{equation}{\label{lambda}}
    \lambda = X^2P_{,XX} +\frac{2}{3}X^3P_{,XXX}\;.
\end{equation}
Here $\Sigma$ is of order $\mathcal{O}(\epsilon)$, but for general
$P(X,\phi^I)$ where $P_{XI} \neq 0$, such as K-inflation or
DBI-inflation, there is no simple relation between $\lambda$ and the
above slow-varying parameters.

\subsection{$\delta N$ Formalism}

In this section we make a brief review of $\delta N$ formalism
\cite{deltaN_starobinsky,9507001,9801017,0411220}, which is proved
to be a powerful technique to calculate the curvature perturbation
in a variety of inflation models, especially in the multi-field
models.

The idea of $\delta N$ formalism is to identify primordial curvature
perturbation $\zeta$ with the perturbation of the local expansion.
Starting from a flat slice at some initial time $t_{\mathrm{i}}$,
the local expansion $N(t,t_{\mathrm{i}},\textbf{x})$ at some final
time $t$ is defined as \be
    N(t,t_{\mathrm{i}},\textbf{x}) = \int_{t_{\mathrm{i}}}^{t} dt'
    H(t',\textbf{x})\;,
    \ee
where $H(t,\textbf{x})$ is the local Hubble expansion rate due to
the perturbations. Then the primordial curvature perturbation can be
expressed as \be
    \zeta(t,\textbf{x}) = N(t,t_{\mathrm{i}},\textbf{x}) - N_0(t,t_{\mathrm{i}}) \equiv \delta N\;,
    \ee
where $N_0(t,t_{\mathrm{i}})$ is the background (unperturbed)
expansion, which is related to the background Hubble expansion rate
$H_0(t)$ as \be
    N_0(t,t_{\mathrm{i}}) = \int_{t_{\mathrm{i}}}^{t}dt'
    H_0(t')\;.
    \ee
If we take $t_{\mathrm{i}}$ at the time of horizon-crossing during
inflation denoted by $t_{\ast}$, then $N(t,t_{\ast},\textbf{x})$
becomes a function of the scalar fields evaluated at
horizon-crossing. Then $\zeta$ can be expanded as
\be{\label{delta_N}}
    \zeta(t,\textbf{x}) = \sum_I
    N_{,I}(t)\delta\phi^I_{\ast}(\textbf{x}) +
    \frac{1}{2}\sum_{I,J}N_{,IJ}(t)\delta\phi^I_{\ast}(\textbf{x})\delta\phi^J_{\ast}(\textbf{x})+\cdots\;,
    \ee
Note that $\zeta(t)$ is just the primordial adiabatic curvature
perturbation if we choose $t$ well after the reheating process.

After going to the momentum space, we have \be
    \zeta(\vk) = N_{,I} \delta \phi^I(\vk) +
    \frac{1}{2}N_{,IJ}[\delta\phi^I\ast\delta\phi^J](\vk)+\cdots\;.
    \ee
The two-point and three-point functions of $\zeta$ can be expressed
in terms of the two and three-point functions of the scalar fields
fluctuations $\delta\phi^I$ as
\begin{align}
    \la \zeta(\vk_1)\zeta(\vk_2) \ra &= N_{,I}N_{,J}\la
    \delta\phi^I(\vk_1)\delta\phi^J(\vk_2)\ra +\cdots\;, {\label{zeta_2pf}}\\
    \la \zeta(\vk_1)\zeta(\vk_2)\zeta(\vk_3) \ra &= N_{,I}N_{,J}N_{,K}\la
    \delta\phi^I(\vk_1)\delta\phi^J(\vk_2)\delta\phi^K(\vk_3)\ra \nonumber\\
    &\qquad\qquad +
    \frac{1}{2}N_{,I}N_{,J}N_{,KL}\la\delta\phi^I(\vk_1)\delta\phi^J(\vk_2)[\delta\phi^K\ast\delta\phi^L](\vk_3)\ra + \mathrm{perms}
    +\cdots\;, {\label{zeta_3pf}}
    \end{align}
    where $\ast$ denotes the convolution
product. In Section \ref{section_2rd} we can see that the two-point
functions for the scalar fields satisfy \be
    \la \delta\phi^I(\vk_1)\delta\phi^J(\vk_2) \ra =
    (2\pi)^3\delta^2(\vk_1+\vk_2)G^{IJ}\frac{2\pi^2}{k_1^3}\Delta_{\star}^2\;,
    \ee
and in Section \ref{section_3pf} we will show that the scalar
three-point functions can be written in the form
\be{\label{A^IJK_def}}
    \la \delta\phi^I(\vk_1)\delta\phi^J(\vk_2)\delta\phi^K(\vk_3)
    \ra =
    (2\pi)^2\delta^3(\vk_1+\vk_2+\vk_3)\frac{4\pi^4}{\prod_ik_i^3}{|\Delta_{\star}^2|}^2\mathcal{A}^{IJK}\;,
    \ee
where $\Delta_{\star}^2$ is the power spectrum of a massless scalar
field in de Sitter space. The principal result of this work is the
momentum-dependent function $\mathcal{A}^{IJK}(k_1,k_2,k_3)$ given
in (\ref{A^IJK}), which contains the information of the amplitude
and shape of the non-Gaussianity.

In order to connect the above analysis with the observations, the
non-Gaussianity measured by the three-point functions must be
expressed in terms of an experimentally relevant parameter. A common
choice is the non-linearity parameter $f_{\mathrm{NL}}$ defined as
\be
    \zeta = \zeta_{\mathrm{g}} +
    \frac{3}{5}f_{\mathrm{NL}}\zeta_{\mathrm{g}}^2\;,
    \ee
which denotes the departure of $\zeta$ from a Gaussian random
variable $\zeta_{\mathrm{g}}$. The power spectrum and bispectrum of
$\zeta$ are defined in terms of the two and three-point functions
respectively as \be
\begin{aligned}
    \la \zeta(\vk_1)\zeta(\vk_2) \ra &= (2\pi)^3\delta^3(\vk_1
    +\vk_2)P_{\zeta}(k_1)\;,\\
    \la \zeta(\vk_1)\zeta(\vk_2)\zeta(\vk_3) \ra &=
    (2\pi)^3\delta^3(\vk_1+\vk_2+\vk_3)B_{\zeta}(k_1,k_2,k_3)\;,
    \end{aligned}
    \ee
then $B_{\zeta}$ is related with $P_{\zeta}$ in terms of
$f_{\mathrm{NL}}$ as\be
    B_{\zeta} = \frac{6}{5}f_{\mathrm{NL}}(k_1,k_2,k_3)[P_{\zeta}(k_1)P_{\zeta}(k_2) + P_{\zeta}(k_2)P_{\zeta}(k_3) +P_{\zeta}(k_3)P_{\zeta}(k_1)]\;.
    \ee
To relate $f_{\mathrm{NL}}$ with the three-point function of
$\zeta$, we write \be
    \la \zeta(\vk_1)\zeta(\vk_2)\zeta(\vk_3) \ra =
    (2\pi)^3\delta^3(\vk_1+\vk_2+\vk_3)\frac{4\pi^4}{\prod_ik_i^3}{|\Delta_{\star}^2|}^2\mathcal{A}_{\zeta}\;,
    \ee
then $f_{\mathrm{NL}}$ can be written as \be
    f_{\mathrm{NL}} =
    \frac{5}{6}\frac{\mathcal{A}_{\zeta}}{\sum_ik_i^3}\;.
    \ee
From (\ref{zeta_2pf}) and (\ref{zeta_3pf}), $\mathcal{A}_{\zeta}$
can also be written as \be
    \mathcal{A}_{\zeta} = N_{,I}N_{,J}N_{,K}\mathcal{A}^{IJK} +
    G^{IK}G^{JL}N_{,I}N_{,J}N_{,KL}\sum_ik_i^3\;.
    \ee
From the above relations, the non-linearity parameter
$f_{\mathrm{NL}}$ can be expressed in terms of $N_{,I}$ and the
momentum-dependent function $\mathcal{A}^{IJK}$, up to leading
orders, as \be{\label{f_NL}}
    f_{\mathrm{NL}} =
    \frac{5}{6}\frac{N_{,I}N_{,J}N_{,K}\mathcal{A}^{IJK}}{(G^{IJ}N_{,I}N_{,J})^2 \sum_ik_i^3}
    + \frac{5}{6}\frac{G^{IK}G^{JL}N_{,I}N_{,J}N_{,KL}}{(G^{IJ}N_{,I}N_{,J})^2} +
    \cdots\;,
    \ee
where `$\cdots$' denotes the remaining cross terms from
(\ref{zeta_3pf}) which we have neglected together with other
higher-order terms. This expression was first derived in
\cite{0504045}.

Furthermore, in order to calculate the primordial power spectrum and
the non-linearity parameter, we need to know the derivatives of the
number of e-folding $N$ with respect to the scalar fields $N_{,I}$,
$N_{,IJ}$, etc. It is easy to show that $dN = -d\ln H/\epsilon$, and
thus \be
    N_{,I} =
    -\sqrt{\frac{P_{,X}}{2}}\frac{\epsilon_I}{\epsilon} +\cdots\;.
    \ee


\section{Linear Perturbations}

\subsection{ADM formalism and the constraint equations}

In single field inflation models, we have two different gauge
choices. One is the gauge where we consider the curvature scalar on
uniform density hypersurfaces, defined by $\delta \phi=0$; the other
is the spatially flat gauge, where we choose the uniform curvature
slicing, and the spatial part of the metric is unperturbed. The
physical degrees of freedom of the perturbations are completely
described by the perturbations of the metric in the first gauge, and
only by the perturbations of the scalar fields in the spatially flat
gauge. In single-field case, both two gauge choices are natural.
However, in the case of multi-field models, the first gauge is no
longer possible, and the only natural choice is the spatially flat
gauge, where the physical degrees of freedom are perturbations of
the scalar fields.

It is very convenient to work in the ADM metric formalism
\begin{equation}
    ds^2 = -N^2 dt^2 +h_{ij}(dx^i+N^idt)(dx^j+N^jdt)\;,
\end{equation}
where $N$ is the lapse function and $N^i$ is the shift vector. The
ADM formalism is convenient because the equations of motion for $N$
and $N^i$ are exactly the energy and momentum constraints which are
quite easy to solve.

Under the ADM metric ansatz, the action becomes{\footnote{Here and
in what follows, the spatial indices $i,j$ are raised and lowered
using $h_{ij}$.}}
\begin{equation}{\label{ADMaction}}
    S = \int dt d^3x \sqrt{h}N\left( \frac{R^{(3)}}{2}+P\right) + \int dtd^3x
    \frac{\sqrt{h}}{2N}\left(E_{ij}E^{ij}-E^2\right)\;,
\end{equation}
where $h=\mathrm{det}h_{ij}$ and the symmetric tensor $E_{ij}$ is
defined as
\begin{equation}
    E_{ij} = \frac{1}{2}\left(\dot{h}_{ij}-\nabla_iN_j-\nabla_jN_i\right)\;,
\end{equation}
and $E \equiv \mathrm{tr}E_{ij} = h^{ij}E_{ij}$. $R^{(3)}$ is the
three-dimensional Ricci curvature which is computed from the metric
$h_{ij}$. The kinetic term $X$ now can be written as
\begin{equation}{\label{X}}
    X = \frac{1}{2N^2}G_{IJ}\pi^I\pi^J -
    \frac{G_{IJ}}{2}\partial^i\phi^I\partial_i\phi^J\;,
\end{equation}
with
\begin{equation}\label{pi}
    \pi^I = \dot{\phi}^I - N^i\partial_i\phi^I\;.
\end{equation}

The equations of motion for $N$ and $N^i$ give the energy constrain
and momentum constraint respectively
\begin{equation}{\label{constraint_N}}
    P -
    \frac{1}{2N^2}\left(E_{ij}E^{ij}-E^2+2P_{,X}G_{IJ}\pi^I\pi^J\right)=0\;,
\end{equation}
\begin{equation}{\label{constraint_Ni}}
    \nabla_j\left( \frac{1}{N}\left(E^j_i - E\delta^j_i\right) \right) =
    \frac{P_{,X}}{N}G_{IJ}\pi^I\partial_i\phi^I\;.
\end{equation}

In spatially flat gauge, we have $h_{ij} =a^2(t)\delta_{ij}$ and
thus $R^{(3)}=0$. The unperturbed flat FRW background corresponds to
$N=1$, $N^i=0$. The scalar fields on the flat hypersurfaces can be
decomposed into \be
    \phi^I(t,\vec{x}) = \phi^I_0(t) + Q^I(t,\vec{x})\;,
    \ee
where $\phi^I_0$ are the spatially homogeneous background values,
and $Q^I$ are the linear perturbations. In what follows, we always
drop the subscript `0' on $\phi^I_0$ and simply identify $\phi^I$ as
the unperturbed background fields. In order to study the scalar
perturbations of the metric and the scalar fields, we may expand $N$
and $N^i$ as \be
\begin{aligned}
    N &= 1+\alpha_1 + \alpha_2 + \cdots\;,\\
    N^i &= \partial^i \beta = \partial^i\left(\beta_1+\beta_2+
    \cdots\right)\;,
\end{aligned}
    \ee
where $\alpha_n,\beta_n$ are of order $\mathcal{O}(Q^n)$. One can
plug the above power expansions into the constrain equations of $N$
and $N^i$ (\ref{constraint_N})-(\ref{constraint_Ni}) to determine
$\alpha_n$ and $\beta_n$. To the first-order of $Q^I$ the solutions
are \cite{0801.1085} \be{\label{alpha_1}}
    \alpha_1 = \frac{P_{,X}}{2H}\dot{\phi}_IQ^I\;,
    \ee
and \be{\label{beta_1}}
    \partial^2\beta_1 = \frac{a^2}{2H}\left[ -\frac{P_{,X}}{c^2_s}\df_I\dot{Q}^I - 2XP_{,XI}Q^I + P_{,I}Q^I + \frac{P_{,X}}{H}\left( \frac{XP_{,X}}{c^2_s} -3H^2\right)\df_IQ^I
\right]\;,
    \ee
where $\partial^2=\delta^{ij}\partial_i\partial_j$. Fortunately, it
turns out that in order to expand the effective action to order
$\mathcal{O}(Q^3)$, in the ADM formalism we do not need to compute
$N$ and $N^i$ to order $\mathcal{O}(Q^3)$, since they must be
multiplied by $\partial L/\partial N$ or $\partial L/\partial N^i$
which vanish due to the constraint equations. Also in the present
case, terms of order $\mathcal{O}(Q^2)$ in $N$ and $N^i$ drop out of
the third-order effective action, and thus
(\ref{alpha_1})-(\ref{beta_1}) are sufficient for our purpose.
Furthermore, it is easy to see that $\alpha_1$ and $\beta_1$ are
both of order $\mathcal{O}(\epsilon^I)$.

\subsection{The second-order theory}{\label{section_2rd}}

In the Appendix \ref{generalexpansion}, the general form of the
expansion of the action to the cubic-order of $Q^I$ has been
developed. From (\ref{S2_origin}), the second-order action can be
written as{\footnote{This expression should be compared, for
example, with eq. (44) of \cite{0506056}, to which it reduces in the
multi-field models with canonical kinetic terms where $P = X-V$.}}
\cite{0801.1085} \be{\label{S2}}
    S^{(2)} = \frac{1}{2}\int dt d^3x a^3 \left[ \left( P_{,X}G_{IJ} + P_{,XX}\df_I\df_J \right)\dot{Q}^I\dot{Q}^J + 2\df_IP_{,XJ} \dot{Q}^IQ^J - P_{,X}G_{IJ}\partial^iQ^I\partial_iQ^J - \mathcal{M}_{IJ}Q^IQ^J\right]\;,
    \ee
where $\mathcal{M}_{IJ}$ is the effective mass-matrix
\be{\label{mass_matrix}}
    \mathcal{M}_{IJ} = -P_{,IJ} + \frac{2XP_{,X}}{H}\df_IP_{,XJ} +
    \frac{XP^3_{,X}}{2H^2}\left(1-\frac{1}{c^2_s}\right)\df_I\df_j -
    \frac{1}{a^3}\frac{\mathrm{d}}{\mathrm{d}t}\left[ \frac{a^3}{2H}P^2_{,X}\left(1+\frac{1}{c_s^2}\right)\df_I\df_J \right]\;.
    \ee
(\ref{S2}) is exact and is valid for arbitrary scalar fields
dynamics.

In general, one may decompose the multi-field perturbations into an
adiabatic mode and $\mathcal{N}-1$ entropy modes, since the
background trajectory specifies a special inflaton direction. Here
in this work, under the assumption that $c_s$ is close to unity, we
may take a different method, to analyze all these perturbations in a
unified and more simpler formalism. To solve the second-order action
(\ref{S2}), we define a symmetric matrix $\alpha_{IJ}$ which
satisfies \be{\label{alphaIJ_def}}
    \alpha_{IK}\alpha_{KJ} = a^2 P_{,X} \mathcal{A}_{IJ}\;,
    \ee
with \be{\label{A_IJ}}
    \mathcal{A}_{IJ} \equiv G_{IJ} +
    \frac{P_{,XX}}{a^2P_{,X}}\phi'_I\phi'_J = G_{IJ} + \frac{u}{\epsilon}\epsilon_{IJ}\;,
    \ee
where a prime denotes the derivative with respect to the conformal
time $\eta$ defined by $d\eta =dt/a$. Here $\mathcal{A}_{IJ}$
denotes the deviation from the standard canonical action, which
reduces to $\mathcal{A}_{IJ} = G_{IJ}$ when considering the standard
kinetic terms, where $u=0$. Then we define the new Mukhanov-type
variables \be
    u^I = \alpha^{IJ}Q^J\;,\qquad Q^I=(\alpha^{-1})^{IJ}u^J\;,
    \ee
or in matrix form $u=\alpha Q$ and $Q=\alpha^{-1}u$. This treatment
is an analogue of that in the single-field case. For standard
single-field inflation models with a canonical kinetic term,
$\a_{IJ}$ reduces to $a$, and $u_I = \a_{IJ}Q_J$ reduces to the
usual rescale relation $u=a\delta\phi$, and thus the matrix
$\alpha_{IJ}$ can be viewed as the generalization of the parameter
$a$ in single-field case, where we may rescale the field as
$u=a\delta \phi$.

In terms of these new variables $u^I$, after changing into conformal
time $\eta$, the second-order action (\ref{S2}) can be rewritten in
a matrix form as \be
\begin{aligned}{\label{S2_matrix_form}}
    S^{(2)} &= \int d\eta d^3x \left[ u'^{\mathrm{T}}u' + u'^{\mathrm{T}}\left( -2\a'\a^{-1} +2a^2\a^{-1}\mathcal{B}\a^{-1} \right)u - a^2 \mathcal{A}^{-1}\partial^iu^{\mathrm{T}}
    \partial_iu\right.\\
    &\qquad\qquad \left.+u^{\mathrm{T}}\left( -2a^2\a^{-1}\a'\a^{-1}\mathcal{B}\a^{-1} - a^4\a^{-1}\mathcal{M}\a^{-1} +\a^{-1}\a'\a'\a^{-1} \right)u \right]\;,
    \end{aligned}
    \ee
where we have defined $\mathcal{B}_{IJ} = \phi'_IP_{,XJ}$ for short.
In deriving (\ref{S2_matrix_form}) we used the fact that
$\a^{-1}\a^{-1} = \mathcal{A}^{-1}/a^2P_{,X}$. The equations of
motion for the scalars can be derived from (\ref{S2_matrix_form})
and written in a compact form \be{\label{2rd_eom}}
\begin{aligned}
    &u'' + \left[ \a^{-1}\a' - \a'\a^{-1} +a^2\a^{-1}(\mathcal{B}-\mathcal{B}^{\mathrm{T}})\a^{-1}
    \right]u'\\
    &\qquad\qquad +\left[ \mathcal{A}^{-1}k^2 + a^2\a^{-1}\left(a^2\mathcal{M} +2\mathcal{H}\mathcal{B} +\mathcal{B}' -(\mathcal{B}-\mathcal{B}^{\mathrm{T}})\a^{-1}\a' \right)\a^{-1} - \a''\a^{-1}
    \right]u=0\;,
    \end{aligned}
    \ee
where $\mathcal{H}=a'/a$. Note that (\ref{2rd_eom}) is exact and no
approximation is made. For standard single field models,
(\ref{2rd_eom}) reduces to the well-known result $u'' +(k^2 +a^2m^2
-a''/a)u=0$ as expected. From (\ref{A_IJ}) it can be seen easily
that in the single-field case, the matrix $\mathcal{A}_{IJ}$ reduces
to a pure number $\mathcal{A}=1/c_s^2$, and thus the first term in
the second line, which is proportional to the wave number $k$,
reduces to $c_s^2 k^2$ as expected.

We would like to make some comments here. From (\ref{A_IJ}) we can
see that the unperturbed background fields velocities $\phi'_I$,
represent a special direction in field space when considering
perturbations of the fields. This is just the so-called adiabatic
direction which has been introduced in \cite{0009131} for
multi-field inflation models with canonical kinetic terms. The
decomposition into adiabatic and entropy modes is equivalent to a
`local' rotation in the space of field perturbations, \be
    \tilde{Q}^n = e^{n}_IQ^I\;,\qquad Q^I = e^I_n \tilde{Q}^n\;,
    \ee
the rotation matrix $e^{n}_I$ is just the projection of a new set of
basis $\{e^n\}$ on $\{\phi_I\}$. The first vector is specified as
\be
    e^1_I = \frac{\df_I}{\sqrt{2X}}\;,
    \ee
which is just the unit local adiabatic vector.
Note that the rotation $e^n_{I}$ is defined locally, and dependents
on the background trajectory. The new field $\tilde{Q}^1$ is just
the adiabatic mode while other modes
$\tilde{Q}^{n}\,(n=2,\cdots,\mathcal{N})$ are entropy modes. After
the local rotation, i.e the decomposition into adiabatic and entropy
modes, it is easy to show that $\mathcal{A}_{IJ} = G_{IJ}
+\left({1}/{c_s^2}-1\right)e^1_Ie^1_J$, and \be
    \mathcal{A}_{IJ}\dot{Q}^I\dot{Q}^J = \frac{1}{c_s^2}\left(\dot{\tilde{Q}}^1 + Z^1_p\tilde{Q}^p\right)\left(\dot{\tilde{Q}}^1 +
    Z^1_q\tilde{Q}^q\right) + \sum_{m\neq1}\left(\dot{\tilde{Q}}^m +
    Z^m_p\tilde{Q}^p\right)\left(\dot{\tilde{Q}}^m + Z^m_q\tilde{Q}^q\right)\;,
    \ee
where we denote $Z^m_n = e^m_I\dot{e}^I_n$ for short.
This decomposition clearly shows that the adiabatic component of the
perturbations $\tilde{Q}^1$ obeys a wave equation where the
propagation speed is the sound speed $c_s$, while the entropy modes
of the perturbations $\tilde{Q}^{n}\,(n=2,\cdots,\mathcal{N})$
propagate with the speed of light $c=1$. This property was first
pointed out in the studying of two-field DBI-inflation
\cite{0709.2666,0709.3299}, but here we see that it turns out to be
a generic feature for general multiple field inflation models.

The decomposition into adiabatic and entropy modes is a powerful
tool to analyse the equations (\ref{2rd_eom}). In general, however,
(\ref{2rd_eom}) is a set of coupled equations and rather complicated
to solve, even in the case of canonical kinetic term. The idea is to
take slow-roll approximations as in the standard slow-roll inflation
models. This is the standard approximation in estimating the
amplitude of the perturbation power spectra, i.e. the two-point
functions. For our purpose, it also supplies a unified treatment
with all these perturbations $Q^I$, rather than decomposing them
into adiabatic and entropy modes.

In single-field models, the mass term can be expressed as a
combination of slow-roll parameters, and thus be neglected to
leading-order in slow-roll approximation. Similarly, the mass-matrix
defined in (\ref{mass_matrix}) can also be written in terms of
slow-varying parameters as \be
\begin{aligned}
    \frac{1}{H^2}\mathcal{M}_{IJ} &=
    -3P_{,X}\left(1+\frac{1}{c_s^2}\right)\epsilon_{IJ} +
    3\tilde{\eta}_{IJ} \\
    &\qquad -
    \frac{12}{\sqrt{P_{,X}}}\epsilon\epsilon_I\left( \frac{P_{,X}}{3c_s^2}\tilde{\epsilon}_I + \tilde{\epsilon}_{I,X}
    \right) -
    \frac{\dot{P_{,X}}}{H}\left(1+\frac{1}{c_s^2}\right)\epsilon_{IJ}
    +
    P_{,X}\left(1+\frac{1}{c_s^2}\right)\epsilon(\epsilon_{IJ}-\eta_{IJ})
    + \frac{2P_{,X}}{c_s^2}s\epsilon_{IJ}\\
    &= -6P_{,X}\epsilon_{IJ} +
    3\tilde{\eta}_{IJ} +\mathcal{O}(\epsilon^2)\;.
    \end{aligned}
    \ee
We also note that $\mathcal{B}_{IJ} = \phi'_I P_{,XJ} \sim
\mathcal{O}(\epsilon)$. Since we assume that $c_s^2$ departs from
unity by a quantity that is first-order in slow-roll, i.e.
$(1/c_s^2-1) \sim \mathcal{O}(\epsilon)$, $\mathcal{A}_{IJ}$ can be
written as \be
    \mathcal{A}_{IJ} = G_{IJ} +\mathcal{O}(\epsilon)\;,
    \ee
and thus $(\mathcal{A}^{-1})_{IJ} = G_{IJ} - \mathcal{O}(\epsilon)$.
Furthermore, from (\ref{alphaIJ_def}) it is easy to show that the
last term in (\ref{2rd_eom}) can be written as \be
    \a''\a^{-1} =
    \frac{a''}{a}(1+\mathcal{O}(\epsilon))\;.
    \ee
Bring all these considerations together, the equations of motion to
the lowest-order of slow-varying parameters reduce to a very simple
form which are just $\mathcal{N}$ decoupled de Sitter-Mukhanov
equations \be{\label{2rd_eom_lowest_order}}
    {u^I}'' +\left( k^2 -\frac{a''}{a} \right)u^I=0\;,
    \ee
where $a''/a = 2/\eta^2 +\mathcal{O}(\epsilon)$.

The solutions to (\ref{2rd_eom_lowest_order}) are standard, we then
find the $Q^I$ two-point functions as \be
    \la Q^I(\eta_1,\textbf{x}_1)Q^J(\eta_2,\textbf{x}_2) \ra
    =G^{IJ}G_{\star}(\eta_1,\textbf{x}_1;\eta_2,\textbf{x}_2)\;,
    \ee
while $G_{\star}$ represents \be
    G_{\star}(\eta_1,\eta_2,\vk) =
    \frac{H^2}{2k^3P_{,X}}\times\left\{
        \begin{aligned}
            &(1+ik\eta_1)(1-ik\eta_2)e^{-ik(\eta_1-\eta_2)}\,, \qquad \eta_1>\eta_2\\
            &(1-ik\eta_1)(1+ik\eta_2)e^{+ik(\eta_1-\eta_2)}\,,
            \qquad \eta_1<\eta_2\;,
        \end{aligned}
        \right.
        \ee
where we have chosen boundary conditions so that $G_{\star}$ behaves
like flat space propagator at very early times, when the
perturbation modes are deep inside the horizon. This corresponds to
the Bunch-Davies vacuum \cite{QF_CS_book}. The power spectra for the
scalar fields on large scales can be read easily \be
    \la Q^I(\vk_1)Q^J(\vk_2) \ra =
    (2\pi)^3\delta^3(\vk_1+\vk_2)G^{IJ}\frac{H^2}{2k_1^3P_{,X}}\;,
    \ee
and therefore the dimensionless power spectra are{\footnote{This
result coincides with (97) and (105) of \cite{0801.1085} in the $c_s
\rightarrow 1$ limit as expected, while in \cite{0801.1085} the
adiabatic and entropy spectra are obtained respectively with an
arbitrary $c_s$.}} \be
    \Delta_{\star}^2 = \frac{H^2}{4\pi^2P_{,X}}\;.
    \ee


\section{Non-linear Perturbations}

\subsection{General form of the third-order action}

In this section we turn to the central calculation of this work, the
third-order piece of the coupled action (\ref{ADMaction}). The
general form of the expansion of the action to the cubic-order of
scalar fields perturbations $Q^I$ is developed in Appendix
\ref{generalexpansion} \be
    S^{(3)} = \int dt d^3x a^3 \left(P^{(3)}+\alpha_1P^{(2)} -\alpha_1\Pi^{(2)}
    +\alpha_1^2\Pi^{(1)} - \alpha_1^3\Pi^{(0)}\right)\;,
    \ee
where $\alpha_1$ is given by (\ref{alpha_1}) and $P^{(n)},\Pi^{(n)}$
can be found in Appendix \ref{generalexpansion}. After a
straightforward but rather tedious calculation, we find \be
\begin{aligned}{\label{S3}}
    S^{(3)} &= \int dt d^3x a^3 \left[ \left( -\frac{P_{,X}^2}{4Hc_s^2}G_{IJ}\df_K - \frac{3\lambda P_{,X}^3}{4H\Sigma^2c_s^4}\df_I\df_J\df_K + \frac{1}{2}G_{IJ}P_{,XK} + \frac{1}{2}\df_I\df_JP_{,XXK} \right)\dot{Q}^I\dot{Q}^JQ^K
    \right.\\
    &\qquad + \left( \frac{3\lambda P_{,X}^3}{4\Sigma H^2 c_s^2}\df_I\df_J\df_K - \frac{P_{,X}}{2H}\df_I\df_JP_{,XK} - \frac{\Sigma c_s^2}{H}\df_I\df_J P_{,XXK} + \frac{1}{2}\df_IP_{,XJK}
    \right)\dot{Q}^IQ^JQ^K\\
    &\qquad + \left( -\frac{P_{,X}\Sigma c_s^2}{4H^2}\df_I\df_JP_{,XK} - \frac{P_{,X}^3}{8H^3}(2\lambda-\Sigma+3H^2)\df_I\df_J\df_K + \frac{P_{,X}^2}{4H^2}\df_I\df_JP_{,K} + \frac{P_{,X}}{4H}P_{,IJ}\df_K \right.\\
    &\qquad \qquad \qquad  \left.+ \frac{\Sigma^2 c_s^4}{2H^2}\df_I\df_JP_{,XXK} - \frac{\Sigma c_s^2}{2H}\df_IP_{,XJK} + \frac{1}{6}P_{,IJK}
    \right)Q^IQ^JQ^K\\
    &\qquad +\left( \frac{1}{2}P_{,XX}G_{IJ}\df_K + \frac{1}{6}P_{,XXX}\df_I\df_J\df_K
    \right)\dot{Q^I}\dot{Q}^J\dot{Q}^K - \frac{P_{,X}^2}{4H}G_{IJ}\df_K\partial^iQ^I\partial_iQ^JQ^K\\
    &\qquad - P_{,X}\dot{Q}_I\partial^i\beta_1\partial_iQ^I +
    \frac{P_{,X}^2}{2H}\df_I\df_JQ^I\partial^i\beta_1\partial_iQ^J -
    P_{,XX}\df_I\df_J\dot{Q}^I\partial^i\beta_1\partial_iQ^J\\
    &\qquad -
    \frac{P_{,XX}}{2}G_{IJ}\df_K\partial^iQ^I\partial_iQ^J\dot{Q}^K
    + \frac{P_{,XX}\Sigma
    c_s^2}{H}\df_I\df_JQ^I\partial^i\beta_1\partial_iQ^J + \frac{P_{,XX}\Sigma
    c_s^2}{2H}G_{IJ}\df_K\partial^iQ^I\partial_iQ^JQ^K\\
    &\qquad \left.- P_{,XI}\df_JQ^I\partial^i\beta_1\partial_iQ^J -
    \frac{1}{2}G_{IJ}P_{,XK}\partial^iQ^I\partial_iQ^JQ^K -
    \frac{P_{,X}}{4H}\df_IQ^I\left(\partial^{ij}\beta_1\partial_{ij}\beta_1 -
    (\partial^i\partial_i\beta_1)^2\right)\right]\;.
    \end{aligned}
    \ee
No approximation of small slow-varying parameters has been made in
deriving (\ref{S3}), and thus it is exact.

\subsection{Slow-varying limit}

In order to proceed, we restrict (\ref{S3}) to the leading-order of
slow-varying parameters. This is because of not only the complexity
of the full cubic-order action (\ref{S3}), but also the
observational constraints. Therefore the third-order action can be
written in a much simpler form{\footnote{This reduces to, for
example, eq. (53) of \cite{0506056} where a canonical kinetic term
was considered, as expected.}} \be{\label{S3-leadingorder}}
\begin{aligned}
    S^{(3)} &= \int dt d^3x a^3 \left[ \left(\frac{1}{2}P_{,XX}G_{IJ}\df_K + \frac{1}{6}P_{,XXX}\df_I\df_J\df_K\right)\dot{Q}^I\dot{Q}^J\dot{Q}^K  \right.\\
    &\qquad +\left(\frac{1}{2}G_{IJ}P_{,XK} - \frac{P_{,X}^2}{4H}G_{IJ}\df_K
    \right)\dot{Q}^I\dot{Q}^JQ^K
    -\frac{P_{,XX}}{2}G_{IJ}\df_K\partial^iQ^I\partial_iQ^J\dot{Q}^K\\
    &\qquad  \left.- \left(
    \frac{P_{,X}^2}{4H}G_{IJ}\df_K + \frac{1}{2}G_{IJ}P_{,XK}\right)\partial^iQ^I\partial_iQ^JQ^K -
    P_{,X}\dot{Q}_I\partial^i\beta_1\partial_iQ^I \right]\;.
    \end{aligned}
    \ee
$\beta_1$ is defined in (\ref{beta_1}) and is also given to the
leading-order by \be{\label{beta_1_leading}}
    \partial^2\beta_1 = -\frac{a^2P_{,X}}{2H}\df_I\dot{Q}^I \;.
    \ee

Now it proves most convenient to take integral by parts to eliminate
the last term containing $\partial_i\beta_1$ in
(\ref{S3-leadingorder}). It is easy to show that
\be{\label{partial_beta}}
\begin{aligned}
    &{\phantom{=}}- \int dt d^3 x
    a^3P_{,X}\dot{Q}_I\partial^i\beta_1\partial_iQ^I\\
    &= \int dt d^3 x \left( -\frac{a^3H}{2}P_{,X}\partial^iQ_I\partial_iQ^I\beta_1 -\frac{a^3}{2}P_{,X}\partial^iQ_I\partial_iQ^I\dot{\beta}_1 + aP_{,X}\dot{Q}_I\partial^2Q^I\beta_1 - \frac{a^3}{2}\dot{P_{,X}}\partial^iQ_I\partial_iQ^I\beta_1
    \right)\\
    &\approx \int dt d^3 x a^3 \left( -\frac{HP_{,X}}{2}\partial^iQ_I\partial_iQ^I\beta_1 -\frac{P_{,X}}{2}\partial^iQ_I\partial_iQ^I\dot{\beta}_1 +
    P_{,X}\dot{Q}_I\partial^i\partial_iQ^I\beta_1 \right)\;,
    \end{aligned}
    \ee
where we have used the fact that $\dot{P_{,X}}/(HP_{,X})$ is order
$\mathcal{O}(\epsilon)$, and we have neglected the last term in the
second line in (\ref{partial_beta}). Now (\ref{beta_1_leading}) can
be used to evaluate $\dot{\beta}_1$ in the last line in
(\ref{partial_beta}). Taking time derivative to both sides in
(\ref{beta_1_leading}) and keeping only the leading-order terms we
have \be{\label{dot_beta}}
    \partial^2\dot{\beta}_1 = -a^2P_{,X}\df_I\dot{Q}^I
    - \frac{a^2P_{,X}}{2H}\df_I\ddot{Q}^I\;.
    \ee
As emphasized in \cite{0506056} that the above equation cannot be
inserted directly into the action (\ref{S3-leadingorder}) since it
contains $\ddot{Q}^I$ and therefore will change the order of the
equations of motion for $Q^I$. However, it is very convenient to
make use of the equations of motion derived from the second-order
theory (\ref{S2}) to eliminate $\ddot{Q}^I$ in (\ref{dot_beta}). To
the lowest-order of slow-varying parameters, the second-order action
(\ref{S2}) reduces to the form \be
    S^{(2)} = \frac{1}{2}\int dt d^3x a^3 P_{,X} G_{IJ} \left( \dot{Q}^I\dot{Q}^J  -
    \partial^iQ^I\partial_iQ^J+ \mathcal{O}(\epsilon)
    \right)\;,
    \ee
and gives \be{\label{ddot_Q}}
    \left.\frac{\delta L}{\delta Q^I}\right|_1 = a^3 P_{,X}\left(-3H
    \dot{Q}_I - \ddot{Q}_I + \partial^i\partial_iQ_I \right) + \mathcal{O}(\epsilon)\;.
    \ee
This vanishes when the perturbations $Q^I$ solve the free Gaussian
theory, but $(\delta L/\delta Q^I)|_1$ will be non-zero when
considering the full theory and of course the third-order
interacting theory. Solving $\ddot{Q}^I$ from (\ref{ddot_Q}) and
inserting them into (\ref{dot_beta}), it follows that
\be{\label{dot_beta_final}}
    \partial^2\dot{\beta}_1 = \frac{a^2P_{,X}}{2}\df_I\dot{Q}^I
    - \frac{P_{,X}}{2H}\df_I\partial^2Q^I + \frac{1}{2Ha}\df^I\left.\frac{\delta L}{\delta
    Q^I}\right|_1+
    \mathcal{O}(\epsilon)\;.
    \ee

Substituting (\ref{dot_beta_final}) into (\ref{partial_beta}) and
performing a lot of integrals by parts, we finally get the
equivalent third-order action \be{\label{S3_final}}
\begin{aligned}
    S^{(3)} &= \int dt d^3x \left\{ \frac{a^3}{2} \left( P_{,XX}G_{IJ}\df_K + \frac{1}{3}P_{,XXX}\df_I\df_J\df_K\right)\dot{Q}^I\dot{Q}^J\dot{Q}^K   \right.\\
    &\qquad + a^3 \left(\frac{1}{2}G_{IJ}P_{,XK} - \frac{P_{,X}^2}{4H}G_{IJ}\df_K \right)\dot{Q}^I\dot{Q}^JQ^K \\
    &\qquad -
    \frac{a^3P_{,XX}}{2}G_{IJ}\df_K\partial^iQ^I\partial_iQ^J\dot{Q}^K
    -\frac{a^3}{2}G_{IJ}P_{,XK}\partial^iQ^I\partial_iQ^JQ^K\\
    &\qquad
    \left. -\frac{a^3P_{,X}^2}{2H}G_{IJ}\df_K\dot{Q}^I\partial^2Q^J\partial^{-2}\dot{Q}^K +\left.\frac{\delta L}{\delta Q^I}\right|_1
    F^I(Q) \right\}\;,
    \end{aligned}
    \ee
with \be
    F^I(Q) = \frac{P_{,X}}{4H}\df^I\left( \partial^{-2}(Q_J\partial^2Q^J) -
    \frac{1}{2}Q_JQ^J
    \right)\;.
    \ee

The last term $\left.\frac{\delta L}{\delta Q^I}\right|_1 F^I(Q)$ in
(\ref{S3_final}) which is proportional to $(\delta L/\delta Q^I)|_1$
can be absorbed by a fields redefinition of $Q^I$ into new fields
$\mathcal{Q}^I$, as in the single-field case. It can be shown easily
that the appropriate fields redefinition is \be
    Q^I = \mathcal{Q}^I - F^I(\mathcal{Q}) = \mathcal{Q}^I + \frac{P_{,X}}{8H}\df^I \mathcal{Q}_J\mathcal{Q}^J - \frac{P_{,X}}{4H}\df^I \partial^{-2}(\mathcal{Q}_J\partial^2\mathcal{Q}^J)\;.
    \ee
Such a fields redefinition where $F^I$ are quadratic in $Q^I$, has
no effect on any of the $\mathcal{O}(Q^3)$ terms in the third-order
action (\ref{S3_final}), and thus we may simply replace $Q^I$ with
$\mathcal{Q}^I$ there. On the other hand, the fields redefinition
indeed modifies the quadratic part of the action, i.e. the Gaussian
action (\ref{S2}), which transforms as \be
    S^{(2)}[Q] \mapsto S^{(2)}[\mathcal{Q}] - \int dt d^3x \left.\frac{\delta L}{\delta
    Q^I}\right|_1 F^I(\mathcal{Q})\;,
    \ee
the second term here cancels the last term in (\ref{S3_final})
exactly, which is proportional to the first-order equations of
motion $\left.\frac{\delta L}{\delta Q^I}\right|_1$.

\section{Calculating the Three Point Function}{\label{section_3pf}}

In this section, we proceed to calculate the scalar fields
three-point functions $\la Q^I(\vk_1) Q^J(\vk_2) Q^K(\vk_3)\ra$,
with the third-order perturbative action (\ref{S3_final}) derived in
the above section. The calculation of the three-point functions is
standard, and thus we simply collect the final results here.

1. Contribution from $\dot{Q}^I\dot{Q}^J\dot{Q}^K$ interaction.

In conformal time $\eta$, this interaction can be written as \be
    \int d\eta d^3 x \frac{a}{H} f_{(1)IJK}{Q'}^I{Q'}^J{Q'}^K\;,
    \ee
with dimensionless coefficient \be
\begin{aligned}
    f_{(1)IJK} &\equiv \frac{H}{2}\left( P_{,XX}G_{IJ}\df_K +
    \frac{1}{3}P_{,XXX}\df_I\df_J\df_K\right)\\
    &= \left( \frac{P_{,X}}{2} \right)^{3/2}\left[ G_{IJ} +\left( \frac{2\lambda}{H^2 \epsilon u} -1 \right)\frac{\epsilon_{IJ}}{\epsilon}
    \right]\frac{u}{\epsilon}\epsilon_k\;,
    \end{aligned}
    \ee
where we have expressed $f_{(1)IJK}$ in terms of slow-varying
parameters defined in Section \ref{section_slow_para}. After a
standard calculation we find the contribution from this term as \be
\begin{aligned}
    &\phantom{==} i(2\pi)^3\delta^3(\vk_1+\vk_2+\vk_3) f_{(1)}^{IJK} \frac{H^6}{P_{,X}^3\prod_i (2k_i^3)} \int_{-\infty}^{0}d\eta
    \frac{1}{-H^2\eta} k_1^2 k_2^2 k_3^2 \eta^3 e^{+iK\eta} + \mathrm{perms} + \mathrm{c.c.}\\
    &= (2\pi)^3\delta(\vk_1+\vk_2+\vk_3) f_{(1)\ast}^{IJK}\frac{H_{\ast}^4}{P_{,X\ast}^3
    \prod_i(2k_i^3)}\frac{4 k_1^2 k_2^2 k_3^2}{K^3} +\mathrm{perms} \;,
    \end{aligned}
    \ee
where $K = k_1 +k_2+k_3$, and an asterisk `$\ast$' denotes that the
corresponding quantities are evaluated at horizon crossing $k= aH$.
Here ``permutation" means total 6 ways of simultaneously rearranging
the indices $I$, $J$ and $K$ and momenta $k_1$, $k_2$ and $k_3$
(i.e. the index `I' is always tied to $k_1$, and so on).

2. Contribution from $\dot{Q}^I\dot{Q}^JQ^K$ interaction. \be
    (2\pi)^3\delta(\vk_1+\vk_2+\vk_3)f_{(2)\ast}^{IJK}\frac{2H^4_{\ast}}{P_{,X\ast}^3
    \prod_i(2k_i^3)}\left( \frac{k_1^2k_2^2}{K} + \frac{k_1^2k_2^2k_3}{K^2} \right) +
    \mathrm{perms}\;,
    \ee
with \be
\begin{aligned}
    f_{(2)IJK} &= \frac{1}{2}G_{IJ}P_{,XK} - \frac{P_{,X}^2}{4Hc_s^2}G_{IJ}\df_K \\
    &= -\left( \frac{P_{,X}}{2}\right)^{3/2}G_{IJ}\left( \frac{3u}{2\epsilon}\tilde{\epsilon}_K + 2\tilde{\epsilon}_K +\frac{6H^2}{P_{,X}}\tilde{\epsilon}_{K,X} +\epsilon_K \right)\;.
    \end{aligned}
    \ee

3. Contribution from $\partial^iQ^I\partial_iQ^J\dot{Q}^K$
interaction. \be
    (2\pi)^3\delta^3(\vk_1+\vk_2+\vk_3)f_{(3)\ast}^{IJK}\frac{2H^4_{\ast}}{P_{,X\ast}^3\prod_i(2k_i^3)}(\vk_1\cdot\vk_2)k_3^2\left( \frac{k_1+k_2}{K^2} + \frac{2k_1k_2}{K^3} +\frac{1}{K}
    \right) +\mathrm{perms}\;,
    \ee
with \be
    f_{(3)IJK} = -\frac{H}{2}P_{,XX}G_{IJ}\df_K = -\left( \frac{P_{,X}}{2}
    \right)^{3/2}\frac{u}{\epsilon}G_{IJ}\epsilon_K\;.
    \ee

4. Contribution from $\partial^iQ^I\partial_iQ^JQ^K$ interaction.
\be
    (2\pi)^3\delta^3(\vk_1+\vk_2+\vk_3)f_{(4)\ast}^{IJK}\frac{2H^4_{\ast}}{P_{,X\ast}^3\prod_i(2k_i^3)}\vk_1\cdot\vk_2\left( -K+\frac{\sum_{i>j}k_ik_j}{K} +\frac{k_1k_2k_3}{K^2} \right)
    + \mathrm{perms}\;,
    \ee
with \be
    f_{(4)IJK} = -\frac{1}{2}G_{IJ}P_{,XK} = \left( \frac{P_{,X}}{2} \right)^{3/2} G_{IJ}\left( \frac{3u}{2\epsilon}\tilde{\epsilon}_K +2\tilde{\epsilon}_K +\frac{6H^2}{P_{,X}}\tilde{\epsilon}_{K,X} \right)\;.
    \ee

5. Contribution from $\dot{Q}^I\partial^2Q^J\partial^{-2}\dot{Q}^K$
interaction. \be
    (2\pi)^3\delta^3(\vk_1+\vk_2+\vk_3)f_{(5)\ast}^{IJK}\frac{2H^4_{\ast}}{P_{,X\ast}^3\prod_i(2k_i^3)}\left( \frac{k_1^2k_2^2}{K} + \frac{k_1^2k_2^3}{K^2} \right)
    + \mathrm{perms}\;,
    \ee
with \be
    f_{(5)IJK} = -\frac{P_{,X}^2}{2H}G_{IJ}\df_K = -2\left( \frac{P_{,X}}{2} \right)^{3/2}G_{IJ}\epsilon_K\;.
    \ee

6. Contribution from the fields redefinition $Q^I \mapsto
\mathcal{Q}^I - F^I(\mathcal{Q})$.

Redefinition $Q^I \mapsto \mathcal{Q}^I + \frac{P_{,X}}{8H}\df^I
\mathcal{Q}_J\mathcal{Q}^J$ gives \be
    (2\pi)^3\delta^3(\vk_1+\vk_2+\vk_3)\frac{H^4_{\ast}}{P_{,X\ast}^3\prod_i(2k_i^3)}\left(\frac{P_{,X\ast}}{2}\right)^{3/2}\epsilon^IG^{JK}k_1^3
    + \mathrm{perms}\;,
    \ee
and redefinition $Q^I \mapsto \mathcal{Q}^I - \frac{P_{,X}}{4H}\df^I
\partial^{-2}(\mathcal{Q}_J\partial^2\mathcal{Q}^J)$ gives
\be
    -(2\pi)^3\delta^3(\vk_1+\vk_2+\vk_3)\frac{H^4_{\ast}}{P_{,X\ast}^3\prod_i(2k_i^3)}\left(\frac{P_{,X\ast}}{2}\right)^{3/2}\epsilon^IG^{JK}k_1k_3^2 +\mathrm{perms}  \;.
    \ee

Bring all these contributions together, after some simplifications,
we find an expression for the scalar fields perturbations
three-point correlation functions \be{\label{QQQ}}
    \la Q^I(\vk_1)Q^J(\vk_2)Q^K(\vk_3)\ra = (2\pi)^3
    \delta^3(\vk_1+\vk_2+\vk_3)
    \frac{H_{\ast}^4}{(2P_{,X})^{3/2}\prod_i(2k_i^2)}\tilde{\mathcal{A}}^{IJK}\;,
    \ee
with \be{\label{A^IJK}}
\begin{aligned}
    \tilde{\mathcal{A}}^{IJK} & =
    G^{IJ}\epsilon^K\frac{u}{\epsilon}\left[ \frac{4k_1^2k_2^2k_3^2}{K^3} - 2(\vk_1\cdot\vk_2)k_3\left( \frac{1}{K} + \frac{k_1+k_2}{K^2} \frac{2k_1k_2}{K^3}\right)
    \right]\\
    &\qquad - G^{IJ}\epsilon^K \left[ 6\frac{k_1^2k_2^2}{K} + 2\frac{k_1^2k_2^2(k_3+2k_2)}{K^2} +k_3k_2^2 -k_3^3
    \right]\\
    &\qquad + G^{IJ}\left( 3\frac{u}{\epsilon} \tilde{\epsilon}^{K}
    +4 \tilde{\epsilon}^{K}  + \tilde{\epsilon}^K_{,X}\frac{12H^2}{P_{,X}}\right) \left[ -\frac{k_1^2k_2^2}{K} - \frac{k_1^2k_2^2k_3}{K^2} + (\vk_1\cdot\vk_2)\left( -K+\frac{\sum_{i>j}k_ik_j}{K} +\frac{k_1k_2k_3}{K^2} \right)
    \right]\\
    &\qquad + \frac{\epsilon^{IJ}}{\epsilon}\epsilon^K\left( \frac{2\lambda}{H^2\epsilon^2} -\frac{u}{\epsilon}
    \right)\frac{4k_1^2k_2^2k_3^2}{K^3} + \mathrm{perms} \;,
    \end{aligned}
    \ee
and $\mathcal{A}_{IJK}$ defined in (\ref{A^IJK_def}) is related to
the above by \be\label{A^IJK_final}
    \mathcal{A}_{IJK} = \frac{1}{4}\sqrt{\frac{P_{,X}}{2}}\tilde{\mathcal{A}}_{IJK}\;.
    \ee
Here the Hubble parameter $H$, effective speed of sound $c_s$,
slow-varying parameters $\epsilon^I$ etc. and $\lambda$ are all
evaluated at the time of horizon-crossing $k \approx aH $. In
general, with an arbitrary $c_s$, the adiabatic mode exits the
horizon at $c_s k=aH$, as addressed by many authors
\cite{0709.2666,0801.1085}. In this paper we assume that $u
=1/c_s^2-1$ is of $\mathcal{O}(\epsilon)$.

The above result (\ref{QQQ})-(\ref{A^IJK_final}) reduces to that of
\cite{0506056}, where the three-point functions of multi-field
models with canonical kinetic term $P=X-V$ have been investigated.
Our result shows the dependence of non-Gaussianity on these
slow-varying parameters. As argued in \cite{0605045}, where a very
general single field model was considered, our results shows that
the final non-Gaussianity is proportional to these small
slow-varying parameters, except a parameter $\lambda$. However, from
the last line in (\ref{A^IJK}) it is easy to see that the large
non-Gaussianities would arise in models with $-
\frac{\lambda}{H^2\epsilon} \gg 1$ during inflation. Note that the
standard choice of kinetic term corresponds to the case $\lambda =
0$. Models with large $u$, i.e. $c_s \ll 1$ may be another source of
large non-Gaussianity, as argued in \cite{0605045}, and it will be
of great interest, we would like to leave this for a future
investigation.


\section{Conclusion}

In this paper, we have calculated the three-point function of
curvature perturbation at the horizon-crossing, which arises from a
general multiple-field inflation model with the action
(\ref{action}). We use $\delta N$ formalism to relate the curvature
perturbation $\zeta$ and the multiple scalar fields fluctuations.
The result is given by (\ref{f_NL}) and
(\ref{QQQ})-(\ref{A^IJK_final}), where the momentum-dependent
amplitude $\mathcal{A}^{IJK}$ is given by
(\ref{A^IJK})-(\ref{A^IJK_final}). This result includes a large
class of inflationary models, and can be viewed as a multiple-field
generalization of previous results considering k-inflation
\cite{k-inflation}, ghost condensation \cite{0312099} and
DBI-inflation \cite{0709.2666,0709.3299,0711.0760}. In the case of
standard canonical kinetic term, our result reduces to the previous
known results \cite{0506056,0804.0574,0802.0588}, where the method
of Lagrangian formalism or the fields equations recently developed
is used. Our formalism would be helpful to analyze the dependence of
non-Gaussianities on the structure of the inflation models, and also
be useful to study the non-Gaussianities in multi-field inflationary
models which will be constructed in the future.

In the presence of many light scalar fields coupled to Einstein
gravity with a non-canonical kinetic term, we define some
slow-varying parameters in order to control the theory and to find
the solutions perturbatively in these small parameters. These
parameters can be seen as a generalization of the standard slow-roll
parameters. However, in this paper we do not ascertain the
conditions under which a particular $P(X,\phi^I)$ would admit such a
slow-varying limit. If this limit breaks done, the theory would
become rather complicated and the calculation would be less clear.
Moreover, as addressed in \cite{0503692,0605045} and etc., in the
presence of a non-canonical kinetic term, the smallness of these
slow-varying parameters do not imply that the inflation itself is
slow-rolling.

In multi-field inflation models, this microphysically-originated
non-Gaussianities produced at the horizon-crossing provide the
initial conditions for the superhorizon evolution of the
non-Gaussianities afterwards \cite{0504045}. Even though the
``initial'' non-Gaussianity is small, after the superhorizon
evolution, the final ``primordial'' non-Gaussianity would be
significantly large, as in the curvaton scenarios
\cite{wands:curvaton,0309033,0110096,0109214,0602285,0607627,0801.0467,0803.2637}.



In this work, we studied the non-Gaussianities which are generated
due to the non-linear relations between curvature and inflaton
perturbations. In general, primordial density perturbations can be
generated not only from the inflaton field(s) during inflation.
There are also other mechanisms which may generate perturbations and
also significant non-Gaussianities. Among such possibilities, for
example, the curvaton mechanism
\cite{wands:curvaton,0309033,0110096,0109214,0602285,0607627,0801.0467,0803.2637},
has been proposed and some of their observational consequences
including issues of the non-Gaussianities have also been
investigated.

As addressed in the introduction, in the multiple-field scenarios,
curvature perturbation can generally evolve after the
horizon-exiting. Detectable non-Gaussianities can be produced when
the curvature perturbation is generated from the entropy
perturbations at the end of inflation \cite{9807278,0207295}, or
during reheating process \cite{0512195,0306006,0511198}. Moreover,
in string inspired inflationary models, other string effects, for
example, cosmic string effect should also be considered
\cite{0204074,0707.0908}.

In this work, we studied the non-linearity of primordial
perturbations measured by three-point functions. Indeed, the
deviation from a Gaussian distribution may also be induced by
higher-order correlation functions. It is interesting to go forward
to find the exact fourth-order action of curvature perturbation and
investigate the non-Gaussianities from the four-point functions and
trispectra \cite{0610210,0610235,0611075,0709.2708,0802.1167}.

Finally, in this work we only considered the case of $c_s \approx
1$, it would be interesting to extend the current result to the case
of an arbitrary, especially a small $c_s$, which would be another
source of large non-Gaussianities. On the other hand, the studies of
multiple DBI-inflation \cite{0709.2666,0709.3299} show that in the
limit of $c_s \ll 1$, the multi-field effects are suppressed by
$c_s$, and thus the multi-field models reduce to a effective
single-field model. We would like to leave this for a future
investigation.



\bigskip
\acknowledgments

I would like to thank Miao Li, Tao Wang, Yi Wang and Wei Xue for
useful discussions and kind help. I am grateful to Miao Li for a
careful reading of the manuscript.



\appendix

\section{General Structure of the Expansion of the Action}{\label{generalexpansion}}

In order to expand the action (\ref{ADMaction}) to the third-order
of $Q^I$, we need the expansions for $P(X,\phi^I)$ and
$E_{ij}E^{ij}-E^2$. Firstly, we expand $P$ and $X$ as \be
\begin{aligned}
    P&=P^{(0)} +P^{(1)}+P^{(2)}+P^{(3)}+\cdots\;,\\
    X&= X^{(0)} +X^{(1)}+X^{(2)}+X^{(3)}+\cdots\;,
    \end{aligned}
    \ee
where $P^{(n)}$ and $X^{(n)}$ are $\mathcal{O}(Q^n)$ pieces of $P$
and $X$ respectively. From (\ref{X})-(\ref{pi}), we have
\be
\begin{aligned}
    X^{(1)} &= \df_I\dot{Q}^I - \frac{P_{,X}X}{H}\df_IQ^I\;,\\
    X^{(2)} &= \frac{1}{2}\dot{Q}_I\dot{Q}^I -
    \df_I\partial^i\beta_1\partial_iQ^I -
    \frac{P_{,X}}{H}\df_I\df_J\dot{Q}^IQ^J +
    \frac{3P_{,X}^2X}{4H^2}\df_I\df_JQ^IQ^J -
    \frac{1}{2a^2}G_{IJ}\partial_iQ^I\partial_iQ^J\;,\\
    X^{(3)} &= -\frac{P_{,X}}{2H}\df_IQ^I\left( \dot{Q}_I\dot{Q}^I-2\df_I\partial^i\beta_1\partial_iQ^I
    \right) + \frac{3P_{,X}^2}{4H^2}\df_I\df_J\df_K\dot{Q}^IQ^JQ^K -
    \frac{P_{,X}^3X}{2H^3}\df_I\df_J\df_KQ^IQ^JQ^K -
    \dot{Q}_I\partial^i\beta_i\partial_iQ^I\;.
    \end{aligned}
\ee
Expanding $P$ to the third-order, we have \be\label{P^n}
\begin{aligned}
    P^{(1)} &= P_{,X}X^{(1)} + P_{,I}Q^I\;,\\
    P^{(2)} &= P_{,X}X^{(2)} +
    \frac{1}{2}P_{,XX}(X^{(1)})^2+\frac{1}{2}P_{,IJ}Q^IQ^J+P_{,XI}X^{(1)}Q^I\;,\\
    P^{(3)} &= P_{,X} X^{(3)} + P_{,XX}X^{(1)}X^{(2)} + P_{,XI}X^{(2)} Q^I \\
        &\qquad\qquad + \frac{1}{6}P_{,XXX}(X^{(1)})^3
    +\frac{1}{2}P_{,XXI}(X^{(1)})^2Q^I+\frac{1}{2}P_{,XIJ}X^{(1)}Q^IQ^J+\frac{1}{6}P_{,IJK}Q^IQ^JQ^K\;.
    \end{aligned}
\ee

The first integral in the action (\ref{ADMaction}) can be expanded
as \be
\begin{aligned}
    S_A &\equiv \int dt d^3x \sqrt{h}NP\\
    &= \int dt d^3 x a^3 (1+\alpha_1
    +\cdots)\left( P^{(0)}+P^{(1)}+P^{(2)}+P^{(3)}
    +\cdots\right)\\
    &=\int dt d^3x a^3\left( P^{(0)} +P^{(1)}+\alpha_1P^{(0)} +P^{2}+\alpha_1P^{(1)} + P^{(3)}+\alpha_1P^{(2)}
    +\cdots
    \right)\;.
    \end{aligned}
    \ee
Therefore the second and the third-order pieces of $S_A$ are
\begin{align}
    S^{(2)}_A &= \int dt d^3x a^3\left( P^{(2)}+\alpha_1P^{(1)}
    \right)\;,\\
    S^{(3)}_A &= \int dt d^3x a^3\left( P^{(3)}+\alpha_1P^{(2)}
\right)\;,
    \end{align}
respectively, where $P^{(n)}$'s are given by (\ref{P^n}).

Now we consider the second integral in (\ref{ADMaction}), which we
denote as $S_B$. First we may expand $E_{ij}E^{ij}-E^2$ as \be
\begin{aligned}{\label{Pi_cap}}
    \Pi &\equiv \frac{1}{2}(E_{ij}E^{ij}-E^2) = \Pi^{(0)} + \Pi^{(1)} + \Pi^{(2)} +\cdots\\
    &= -3H^2 +
    \frac{2H}{a^2}\partial^2\beta_1+\left(\frac{1}{2a^4}\partial_i\partial_j\beta_1\partial_i\partial_j\beta_1
    - \frac{1}{2a^4}(\partial^2\beta_1)^2\right)+\cdots\;,
    \end{aligned}
    \ee
then \be
\begin{aligned}
    S_B &=\int dt d^3x \frac{a^3}{N}\Pi\\
    &= \int dt d^3 x
    a^3\left[ \Pi^{(0)} + (\Pi^{(1)}-\alpha_1\Pi^{(0)})+(\Pi^{(2)}-\alpha_1\Pi^{(1)} +\alpha_1^2\Pi^{(0)}) + (-\alpha_1\Pi^{(2)} +\alpha_1^2\Pi^{(1)}-\alpha_1^3\Pi^{(0)})+\cdots\right]\;,
    \end{aligned}
    \ee
The second and third-order pieces of the second integral in the
coupled action (\ref{ADMaction}) now read
\begin{align}
    S_B^{(2)} &= \int dt d^3x a^3(\Pi^{(2)}-\alpha_1\Pi^{(1)}
    +\alpha_1^2\Pi^{(0)})\;,\\
    S_B^{(3)} &= \int dt d^3x a^3(-\alpha_1\Pi^{(2)}
    +\alpha_1^2\Pi^{(1)}-\alpha_1^3\Pi^{(0)})\;,
    \end{align}
where $\Pi^{(n)}$'s are given by (\ref{Pi_cap}).

Finally, the second and third-order pieces of the effective action
are
\begin{align}
    S^{(2)} &= S_A^{(2)}+S_B^{(2)} = \int dt d^3x a^3 \left(P^{(2)}+\alpha_1P^{(1)} + \Pi^{(2)}- \alpha_1\Pi^{(1)}
    +\alpha_1^2\Pi^{(0)}\right)\;,{\label{S2_origin}}\\
    S^{(3)} &= S_A^{(3)}+S_B^{(3)} = \int dt d^3x a^3 \left(P^{(3)}+\alpha_1P^{(2)} -\alpha_1\Pi^{(2)}
    +\alpha_1^2\Pi^{(1)} - \alpha_1^3\Pi^{(0)}\right)\;.{\label{S3_origin}}
\end{align}




\end{document}